# Nonlinear optics and optical limiting properties of multifunctional fullerenol/polymer composite

**Hendry Izaac Elim**[*], **Wei Ji** and **Gwee Chen Meng**
Department of Physics, National University of Singapore, 3 Science Drive 3, Singapore 117543, Singapore

**Jianying Ouyang** and **Suat Hong Goh**
Department of Chemistry, National University of Singapore, 3 Science Drive 3, Singapore 117543, Singapore

The nonlinear optics and optical limiting properties of materials based on multifunctional fullerenol and poly(styrene-co-4-vinylpyridine) matrix were studied using 7 ns pulses of nanosecond laser operating at 532 nm wavelength. The observed imaginary and real parts of third order susceptibility of the fullerenol/polymer composite are found to be lower than that of its parent $C_{60}$. The optical limiting performances of fullerenol and fullerenol incorporated with poly(styrene-co-4-vinylpyridine) have been proved to be poorer than that of $C_{60}$ due to their higher limiting thresholds. Concentration dependence of poly(styrene-co-4-vinylpyridine) containing 32 mol% has been mainly contributed to the optical limiting performance of fullerenol.

[*] email address: *scip1109@nus.edu.sg*

## 1.Introduction

The nonlinear optical and optical limiting properties of $C_{60}$ have been widely known in resent years due to their promising use as optical limiters for eye and sensor protection from high intensity beams.[1,2] Another important application of this material is its use in optical switches and optically bistable devices.[3] In addition, some investigations about the application of a single-$C_{60}$ transistor, the strong magnetic signals found in rhombohedral $C_{60}$ polymer (Rh-$C_{60}$), and the study of synthetic route to the $C_{60}H_{30}$ polycyclic aromatic hydrocarbon and its laser-induced conversion into fullerene-$C_{60}$ have also received considerable attention.[4-6]

The enhancement of the nonlinear optical and optical limiting properties of $C_{60}$ is a great challenge. A possibility to overcome this is by connecting and interacting the $C_{60}$ with complementary functional groups of a polymer.[7,8] However, the properties of $C_{60}$ may have changed upon functionalization and after mixing with the polymer. Sun and his co-workers have investigated the optical limiting properties of a series of methano-$C_{60}$ derivatives, in which mono-functional $C_{60}$ derivatives show similar optical limiting responses as that of parent $C_{60}$ while the multifunctional derivatives have poorer performance.[9-11] Here we report a comprehensive study on the nonlinear optics and the optical limiting properties of multifunctional fullerenol/polymer composite.

## 2.Preparation of sample

The sample of [60]Fullerene ($C_{60}$) (99.9%) was obtained from Peking University, China. The main procedure to synthesize and characterize the multi-functional fullerenol (Fol) was based on a method used by Chiang, *et al.*[12] We found Fol consist of 10 to 12 hydroxy addends. Toluene and N,N-dimethylformamide (DMF) of AR grade were purchased from Fisher Scientific Company, USA and used as received. Poly(styrene-co-4-vinylpyridine) (PSVPy) sample containing 32 mol% of vinylpyridine, denoted as PSVPy32, was prepared by free radical copolymerization. Fol was dissolved in DMF, into which an appropriate amount of PSVPy32 or polystyrene was added.[13]

Figure 1 shows the ground-state absorption measurement of $C_{60}$ in toluene, Fol in DMF, and Fol/PSVPy32 in DMF, respectively. The spectra indicate that the initial transmittance (T = 70%) of all samples have the same at 532 nm wavelength.

**3.Theoretical and Experimental Descriptions**

The nonlinear optical and optical limiting properties of materials based on multifunctional fullerenol (Fol) and poly(styrene-co-4-vinylpyridine) matrix were studied using 7 ns pulses of nanosecond laser (Q-switched frequency-doubled Nd:YAG) operating at 532 nm wavelength.

The measurements of nonlinear absorption coefficient and nonlinear refractive index were conducted using open and closed aperture z-scan techniques, respectively.[14] The technique was based on the variation of transmitted radiation intensity by alteration of the geometrical parameters of the interaction region. The experimental data were recorded by gradually moving a sample along the axis of propagation ($z$) of a focused Gaussian beam through its focal plane and measuring the transmission of the sample for each $z$ position (Fig. 2). As the sample experiences different electric field strengths at different $z$ position, the recording of the transmission as a function of the $z$ coordinate provides accurate information about the nonlinear effects present.

The open aperture z-scan was performed by assuming a nonlinear absorption effect such that the total absorptivity is given by an expression of the type $\alpha = \alpha_0 + \beta I$.

Additionally, where $\alpha_0$ and $\beta$ are the linear and nonlinear absorption coefficients, respectively. The laser beam used is assumed to be Gaussian in space and time. The normalized transmission $T_n(z)$ for an open aperture $z$-scan can be described by the equation[14]

$$T_n(z) = \frac{C}{\sqrt{\pi}\, q_0} \int_{-\infty}^{\infty} \ln(1 + q_0 e^{-t^2})\, dt \,, \tag{1}$$

where $C$ is a normalization constant and $q_0$ is given by the equation

$$q_0 = \frac{\beta I_0(t) L_{eff}}{(1 + z^2 / z_0^2)}, \tag{2}$$

where $I_0(t)$ is the on-axis instantaneous intensity of the laser beam at the focus, $L_{eff} = [1-\exp(-\alpha_0 L)]/\alpha_0$ is the effective thickness, $L$ is the pathlength of the sample, and $z_0$ is the diffraction length of the laser beam, defined by $z_0 = \pi\omega_0^2 / \lambda$, with $\omega_0$ denoting the beam waist and $\lambda$ the laser wavelength. The normalization of the transmission is essential in order to exclude linear absorption effects. The nonlinear absorption coefficient ($\beta$) is then be determined by fitting Eq. (1) to the experimental data of $T_n(z)$.

The closed aperture z-scan was conducted based on a consideration that a laser beam Gaussian in space and time, traveling in a nonlinear medium which has an intensity-dependent total refractive index given by $n = n_0 + n_2 I$. In the case of negative nonlinearity the normalized transmission in a closed aperture z-scan shows a prefocal transmission maximum (peak), followed by a postfocal transmission minimum (valley). For a thin medium ($L / n_0 z_0 < 1$, where $n_0$ is the linear refractive index) and under the assumption that $q_0(z, t = 0) \le 1$ and Re($\chi^{(3)}$) > Im($\chi^{(3)}$), one can estimate the effect of nonlinear refraction by dividing the data of a closed aperture z-scan by that of an open aperture z-scan, both z-scans being performed at the same incident intensity, one can then determine from the resultant curve the difference between the peak and the valley of the normalized transmittance ($\Delta T_{p-v}$). The nonlinear refractive index parameter $n_2$ is finally estimated by the following equation[14]

$$n_2 = \frac{\Delta T_{p-v} \lambda \alpha}{0.812 \pi I_0 (1-S)^{0.25} (1-e^{-\alpha L})} \tag{3}$$

where $S = 1 - \exp(-2r_a^2 / \omega_a^2)$ is the linear aperture transmission with $r_a$ and $\omega_a$ being the aperture and the beam radii respectively and $\alpha$ is the total maximum absorption of the sample. From the open aperture results, one can obtain the calculation of $\alpha$ and the nonlinear absorption parameter $\beta$ , respectively.

The imaginary and real parts of the third-order nonlinear optical susceptibility $\left(\chi^{(3)}\right)$ were evaluated from the nonlinear absorption $(\beta)$ and refractive index $\left(n_2\right)$ coefficients, respectively. The relations are defined as follows

$$\mathrm{Im}\left(\chi^{(3)}\right) = n_0^2 \varepsilon_0 c \lambda \beta / 2\pi , \tag{4}$$

and

$$\mathrm{Re}\left(\chi^{(3)}\right)(\mathrm{esu}) = \frac{n_0^2}{0.0395} n_2 \left(\mathrm{cm}^2/\mathrm{W}\right), \tag{5}$$

where $n_0$ is the linear refraction index, $\varepsilon_0$ is the vacuum permitivity, and $c$ is the light velocity in vacuum.

The optical limiting measurement of multifunctional fullerenol (Fol)/polymer composites was conducted by measuring fluence-dependent transmittance of the sample in the focal point of the optical limiting experimental setup (Fig. 2). By plotting the normalized transmittance against the input fluence, both the limiting threshold (defined as the input fluence at which the transmittance falls to 50% of the linear transmittance) and optical damage can be readily observed.[15]

**4.Results and Discussion**

To assess the potential application of this multifunctional fullerenol/polymer composite, nonlinear optical (NLO) measurements were conducted. Figure 3 displays the open aperture results of 10-mm-thick diluted and 1-mm-thick concentrated samples of $C_{60}$, Fol, and Fol/PSVPy (Folp), respectively. The data shows that all of the samples have concentration dependence. One obtains that higher concentration of the samples give better NLO properties. The nonlinear absorption coefficient ($\beta$) of Fol and Fol/PSVPy32 (Folp) has been found to be poorer than that of its parent $C_{60}$. In addition, $\beta$ of concentrated Fol/PSVPy32 was found to be a little bit higher than that of Fol. This indicates a high concentration of poly(styrene-co-4-vinylpyridine) (PSVPy) has marginally contributed to the increase of the NLO property of the multifunctional fullerenol/polymer composite due to the distortion caused by interaction between PSVPy and the multifunctional fullerenol.

Figures 4(a) and 4(b) show open aperture z-scans performed with two different input energies (10 and 30 μJ) on diluted and concentrated samples of $C_{60}$, Fol, and Fol/PSVPy32 (Folp), respectively. By fitting the experimental data using Eq.(1), we found that in the case of higher input energy (30 μJ), the nonlinear absorption for all of the samples becomes slightly reduced due to a reverse saturable absorption (RSA). According to the calculated $\beta$ (in Table 1), one finds that the nonlinear absorption coefficients ($\beta$) of Fol and Fol/PSVPy32 (Folp) are lower than that of their parent $C_{60}$.

The weaker nonlinear optical response of Fol and Fol/PSVPy may be ascribed to the disturbance of $\pi$-electronic system of parent $C_{60}$ molecule due to multifunctionalization.

Figure 5 shows the closed aperture z-scans of $C_{60}$, Fol and Fol/PSVPy32 (Folp), respectively. The signals represent a strong concentration dependence of nonlinear refractive index. The concentrated samples provide a better $n_2$ than that of the diluted samples. Compared to Fol-c, the $n_2$ of Folp-c has been improved due to the higher polymer concentration.

Table 1 shows that the observed imaginary and real parts of third order susceptibility of the fullerenol/polymer composite are found to be lower than that of its parent $C_{60}$. Therefore, the multifunctional fullerenol/polymer composite are not effective in increasing the NLO properties of $C_{60}$.

The measurements of optical limiting effects in the DMF solution of multifunctional fullerenol/polymer composites are depicted in Fig. 6. Figure 6 shows the relation between normalized transmittance of diluted and concentrated samples of Fol, Fol/PSVPy32 and $C_{60}$ measured by optical limiting technique with laser light of $\lambda$ = 532 nm and input fluence. When the input fluence is small, the relation is exactly a linear one. However, when the input fluence is gradually increased, the measured normalized transmittance deviates from linearity. It is obviously as shown on table 2 that both Fol and Fol/PSVPy32 limiting thresholds ($F_{th}$) are higher than that of $C_{60}$ which indicates that their optical limiting performances are poorer than their parent. The best limiting threshold of all the samples is obtained from concentrated $C_{60}$ (~2.3 J/cm$^2$).

The same nonlinear characteristics of these multifunctional fullerenol/polymer composite performed by Z-scans and optical limiting measurements indicate that higher absorptive and refractive nonlinearities give better optical limiting performance.

Figure 7 shows the relation between polymer concentration and pyridine/hydroxyl molar ratio on Fol/PSVPy32 and the optical limiting performance of the samples. Evidently, the performance of Fol/PSVPy32-$c_2$ has been improved due to its higher polymer concentration and pyridine/hydroxyl molar ratio than that in Fol/PSVPy32-$d_1$ and Fol/PSVPy32-$c_1$, respectively. Based on a same molar ratio of pyridine/hydroxyl, the improvement effect of poly(styrene-co-4-vinylpyridine) sample containing 32 mol% (PSVPy32) on the optical limiting performance is more prominent in concentrated Fol solution than in dilute Fol solution. In addition, it is shown in Fig. 8 that the main contribution is from polymer concentration. The higher polymer concentration in Fol/PSVPy32-$c_2$ compared to that in Fol/PSVPy32-$d_1$ and Fol/PSVPy32-$c_1$, resulted in better optical limiting properties. On the other hand, Fol/PSVPy32-$d_1$ which has higher pyridine/hydroxyl molar ratio than that in Fol/PSVPy32-$c_1$ shows slightly better optical limiting performance even its polymer and fullerene concentrations are lower than that of Fol/PSVPy32-$c_1$. Therefore, the higher pyridine/hydroxil molar ratio has only slightly contributed to the optical limiting performance.

Figure 9 displays the “inert” polystyrene at various concentrations on Fol/PS does not significantly change the optical limiting performance of Fol solution, while the “active” PSVPy32 improves it. This is probably due to a very small interaction between polystyrene and multifunctional fullerenol.

## 5.Conclusions

The Z-scan and optical limiting measurement techniques have been used to study the nonlinear optical and optical limiting properties of multifunctional fullerenol/polymer composite. The Fol/PSVPy32 posses slightly better NLO as well as its optical limiting properties than that of Fol/PS. However, to make an excellent optical limiter, the materials of multifunctional fullerenol/polymer composite should be modified to sustain a comparable optical limiting performance as that of $C_{60}$.

**Table 1.** Ground-state absorption measured using UV-Vis spectrometer and nonlinear optical parameters of diluted and concentrated samples of $C_{60}$, Fol and Fol/PSVPy measured at the same I = 69.7 $MW/cm^2$.

| Sample | Solvent | Concentration (M) | Transmittance (%) | ε at λ = 532 nm ($M \cdot cm^{-1}$) | β (cm/GW) | $n_2$ (x $10^{-12}$ $cm^2/W$) | Im($\chi^{(3)}$) (x $10^{-11}$ esu) | Re($\chi^{(3)}$) (x $10^{-11}$ esu) |
|---|---|---|---|---|---|---|---|---|
| $C_{60}$-d | Toluene | 1.65x$10^{-4}$ | 70.1 | 940 | 46 | -0.43 | 1.7 | -2.5 |
| $C_{60}$-c | Toluene | 1.65 x$10^{-3}$ | 70.1 | 941 | 165 | -2.06 | 6.0 | -11.7 |
| Fol-d | DMF | 1.17 x$10^{-4}$ | 69.3 | 1360 | 13 | -0.21 | 0.47 | -1.2 |
| Fol-c | DMF | 1.42 x$10^{-3}$ | 70.3 | 1080 | 82.1 | -1.08 | 3.0 | -6.2 |
| Fol/PSVPy32-d | DMF | 1.17 x$10^{-4}$ | 69.6 | 1344 | 12.7 | -0.19 | 0.46 | -1.1 |
| Fol/PSVPy32-c | DMF | 1.42 x$10^{-3}$ | 69.5 | 1112 | 90.3 | -1.19 | 3.3 | -6.8 |

**Table 2.** Optical limiting parameter of $C_{60}$, Fol, Fol/PS and Fol/PSVPy32 measured using nanosecond laser at the same λ = 532 nm and T = 70%.

| Sample | Solvent | Fullerene concentration (M) | Polymer Concentration (mg/ml) | Pyridine/Hydroxyl Molar Ratio | Limiting threshold (J/cm$^2$) |
|---|---|---|---|---|---|
| $C_{60}$-d | Toluene | $1.65x10^{-4}$ | - | - | >5.2 |
| $C_{60}$-c | Toluene | $1.65\ x10^{-3}$ | - | - | 2.3 |
| Fol-d | DMF | $1.17\ x10^{-4}$ | - | - | >10 |
| Fol-c | DMF | $1.42\ x10^{-3}$ | - | - | >10 |
| Fol/PS-$d_1$ | DMF | $1.17x10^{-4}$ | 1.06 | - | >10 |
| Fol/PS-$c_1$ | DMF | $1.42x10^{-3}$ | 1.06 | - | >10 |
| Fol/PS-$c_2$ | DMF | $1.42x10^{-3}$ | 5.37 | - | >10 |
| Fol/PSVPy32-$d_1$ | DMF | $1.17x10^{-4}$ | 0.45 | 0.99/1 | >8 |
| Fol/PSVPy32-$c_1$ | DMF | $1.42x10^{-3}$ | 1.01 | 0.18/1 | >9 |
| Fol/PSVPy32-$c_2$ | DMF | $1.42x10^{-3}$ | 5.05 | 0.91/1 | >6 |

**A list of Figure Captions:**

**Fig. 1.** The ground -state absorption spectra of (1): $C_{60}$ in toluene, (2): Fol in DMF, (3): Fol/PSVPy32 in DMF. The inset is the multifunctional fullerenol (Fol) structure without PSVPy32 and polystyrene.

**Fig. 2.** The experimental setup of z-scans and optical limiting techniques used in this investigation

**Fig. 3.** Open aperture z-scans conducted with 532 nm, 7 ns laser pulses on 10-mm-thick diluted and 1-mm-thick concentrated samples of $C_{60}$, Fol and Fol/PSVPy32 (Folp), respectively. The filled and unfilled circles, squares and triangles are the diluted (d) and concentrated (c) samples, respectively. The solid curves are the fitting curve for diluted samples while the dashed curves are for the concentrated samples. The input energy used is 10 μJ with the beam waist, $\omega_0 = 35$ μm (I = 69.7 MW/cm$^2$).

**Fig. 4.** Open aperture z-scans performed with two different input energies on: (a) 10-mm-thick diluted and (b) 1-mm-thick concentrated samples of $C_{60}$, Fol and Fol/PSVPy32 (Folp), respectively. The filled and unfilled circles, squares and triangles are the observed data at 10 and 30 μJ, respectively. The solid curves are the fitting curve for the samples at 10 μJ, while the dashed curves are for the same samples at 30 μJ.

**Fig. 5.** Closed aperture z-scans conducted with the same I = 69.7 MW/cm$^2$ and $\lambda = 532$ nm on 10-mm-thick diluted and 1-mm-thick concentrated samples of $C_{60}$, Fol and Fol/PSVPy32 (Folp), respectively. The filled and unfilled circles, squares and triangles

are the diluted (d) and concentrated (c) samples, respectively. The solid curves are the fitting curve for the samples.

**Fig. 6.** Optical limiting curves showing the performances of $C_{60}$ in toluene (filled circles and squares), fullerenol (Fol) in DMF (open circles and squares) and Fol/PSVPy32 (filled triangles and inverted triangles) in DMF with respect to the focus as a function of input fluence ($\omega_0$ = 35 μm, $\tau$ = 7 ns (FWHM), and T = 70 %).

**Fig. 7.** Nonlinear transmission as a function of input fluence of fullerenol in DMF (Fol-c) (filled circles), Fol/PSVPy32-$d_1$ in DMF (open circles), Fol/PSVPy32-$c_1$ in DMF (open triangles) and Fol/PSVPy32-$c_2$ in DMF (open squares) measured with nanosecond laser pulses of 532 nm wavelength. All the samples have the same initial transmission ($T \approx 70$ %).

**Fig. 8.** Optical limiting responses showing the performances of Follerenol in DMF (Fol-c) (filled circles), Fol/PSVPy32-$d_1$ in DMF (filled squares), Fol/PSVPy32-$c_1$ in DMF (filled triangles) and Fol/PSVPy32-$c_2$ in DMF (filled inverted triangles) measured with nanosecond laser at $\lambda$ = 532 nm. Shown in the inset is a plot of the output light intensities for the samples vs those for $C_{60}$ at the same input fluence and the same linear transmittance of T = 70 % at the wavelength.

**Fig. 9.** Nonlinear transmission as a function of input fluence of Fol/PSVPy32-$c_2$ in DMF (filled squares), Fol/PS-$d_1$ in DMF (open circles), Fol/PS-$c_1$ in DMF (open triangles) and Fol/PS-$c_2$ in DMF (open squares) measured with nanosecond laser pulses of 532 nm wavelength. All the samples have the same initial transmission ($T \approx 70$ %).

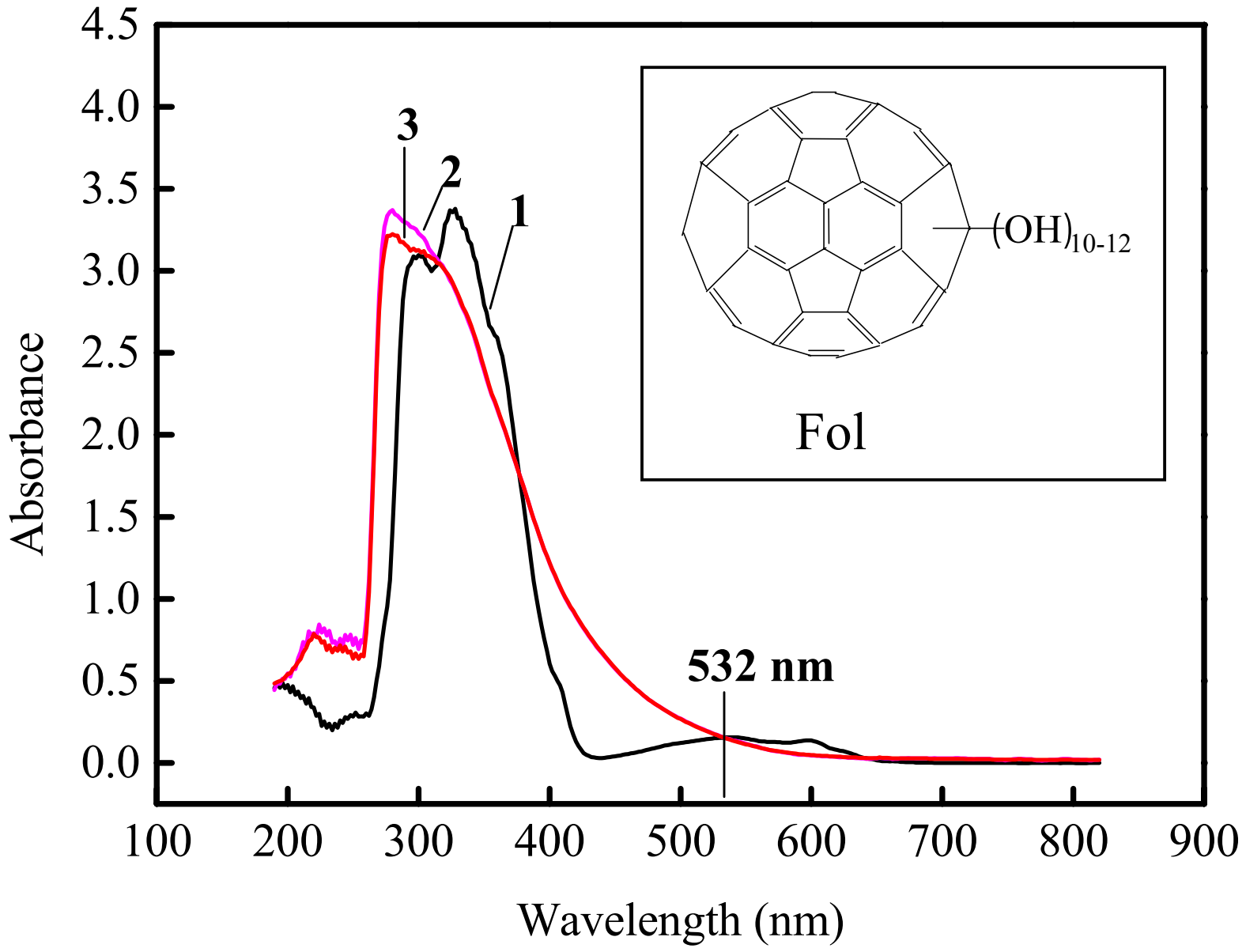

Fig. 1.

**"Nonlinear optics and optical limiting properties of multifunctional fullerenol/polymer composite"**
(Hendry Izaac Elim, *et al*)

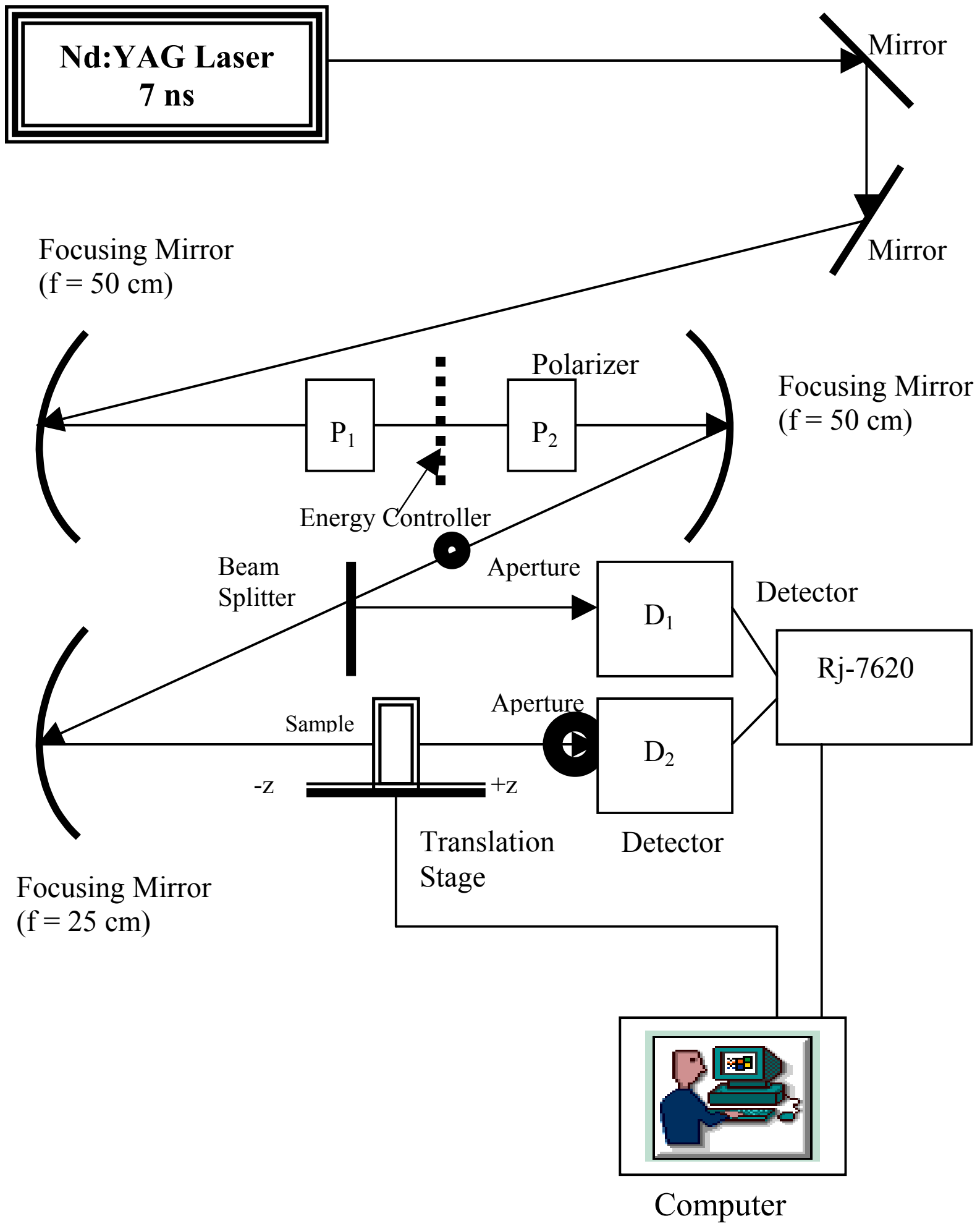

Fig. 2.

**"Nonlinear optics and optical limiting properties of multifunctional fullerenol/polymer composite"**

(Hendry Izaac Elim, *et al*)

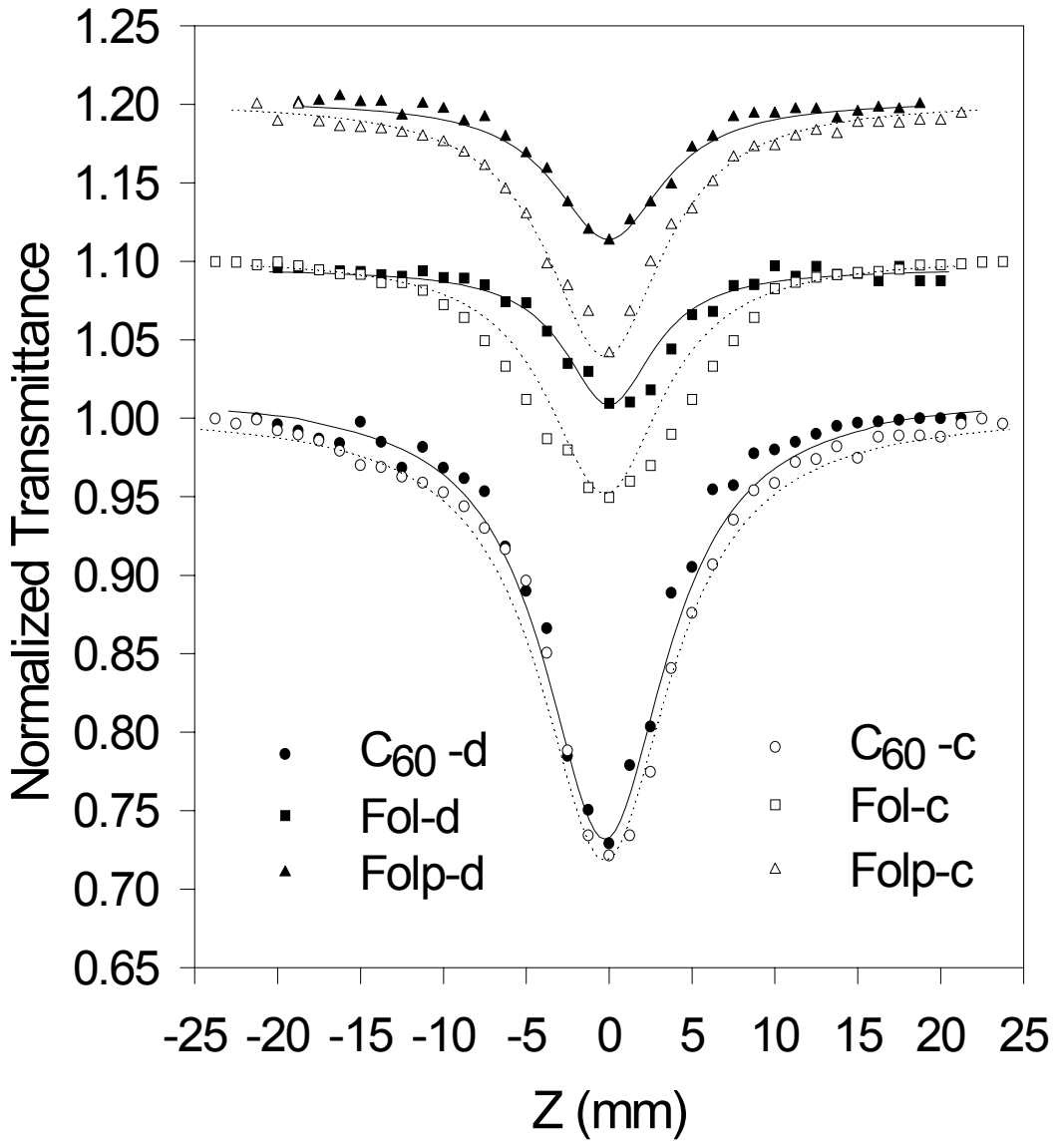

Fig. 3.

**"Nonlinear optics and optical limiting properties of multifunctional fullerenol/polymer composite"**
(Hendry Izaac Elim, *et al*)

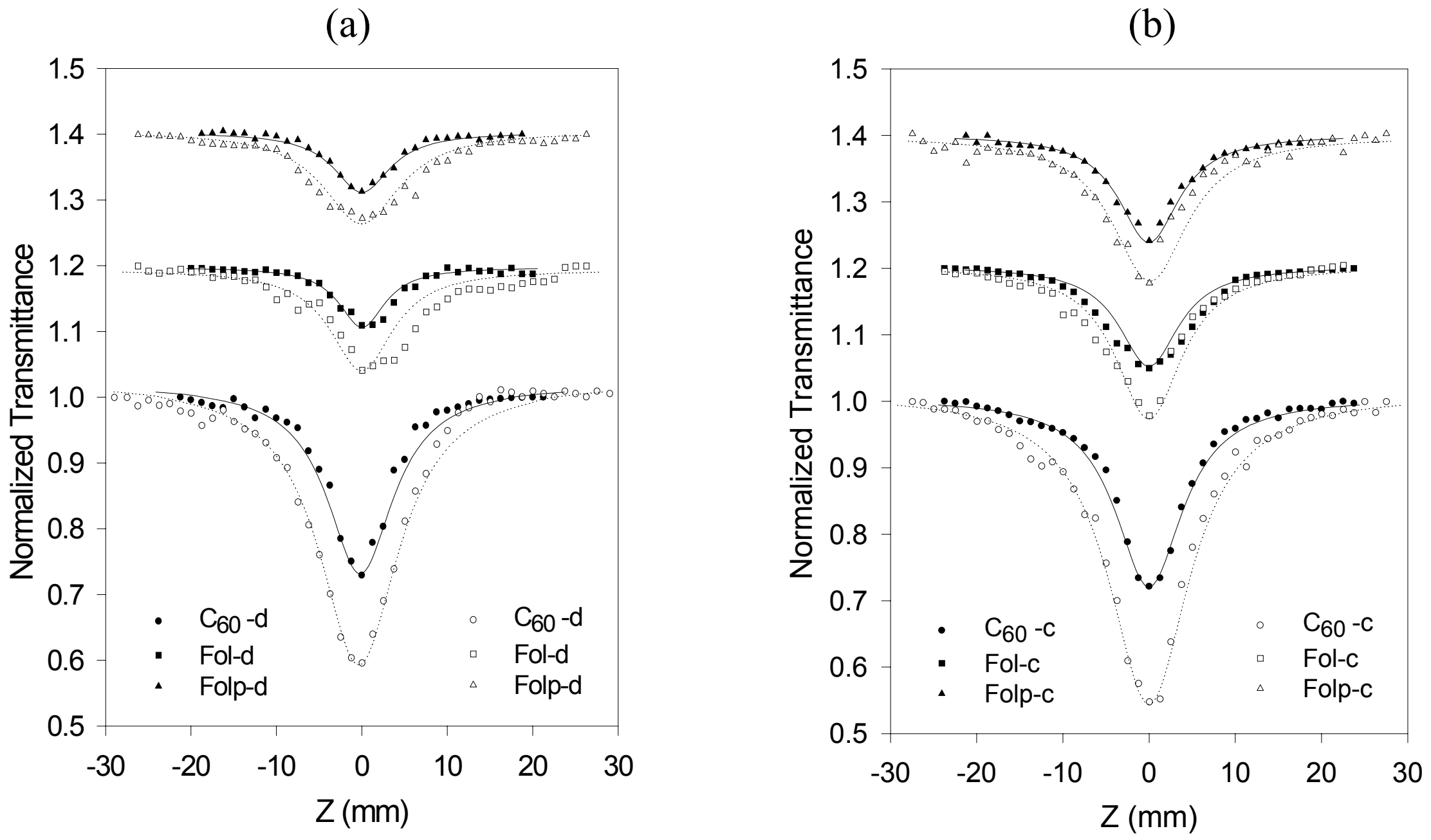

Fig. 4.

**"Nonlinear optics and optical limiting properties of multifunctional fullerenol/polymer composite"**

(Hendry Izaac Elim, *et al*)

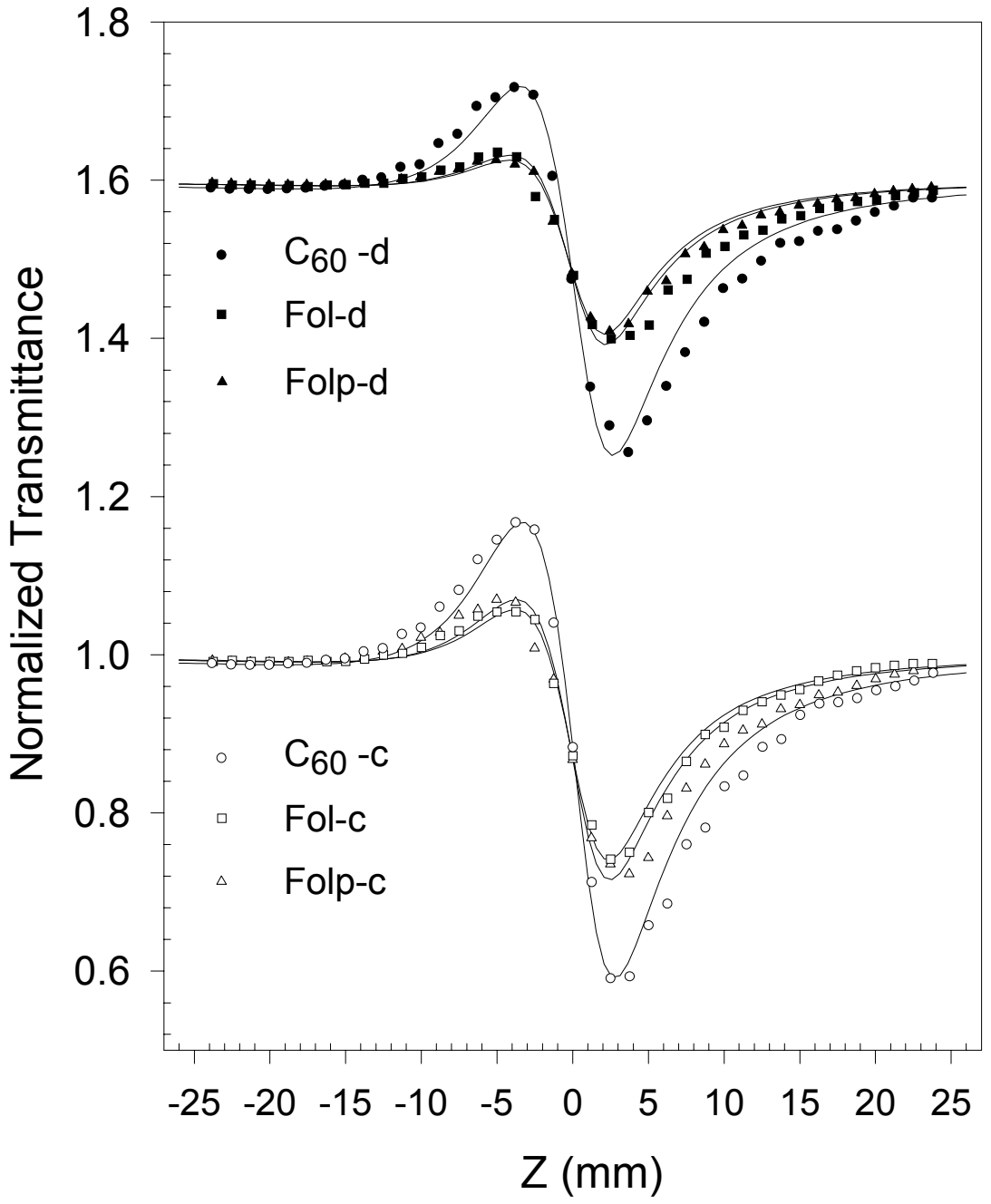

Fig. 5.

**"Nonlinear optics and optical limiting properties of multifunctional fullerenol/polymer composite"**

(Hendry Izaac Elim, *et al*)

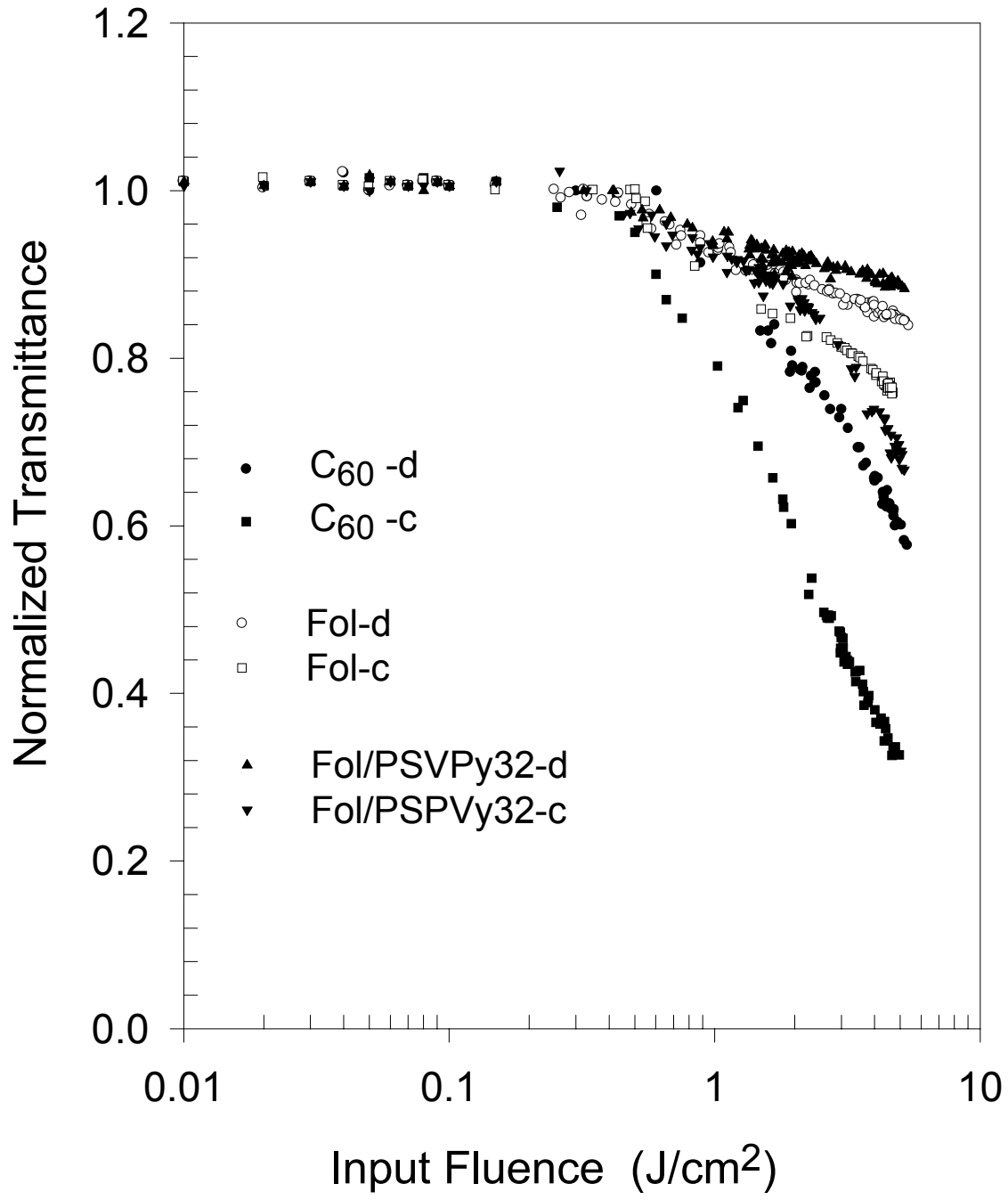

Fig. 6.

**"Nonlinear optics and optical limiting properties of multifunctional fullerenol/polymer composite"**

(Hendry Izaac Elim, *et al*)

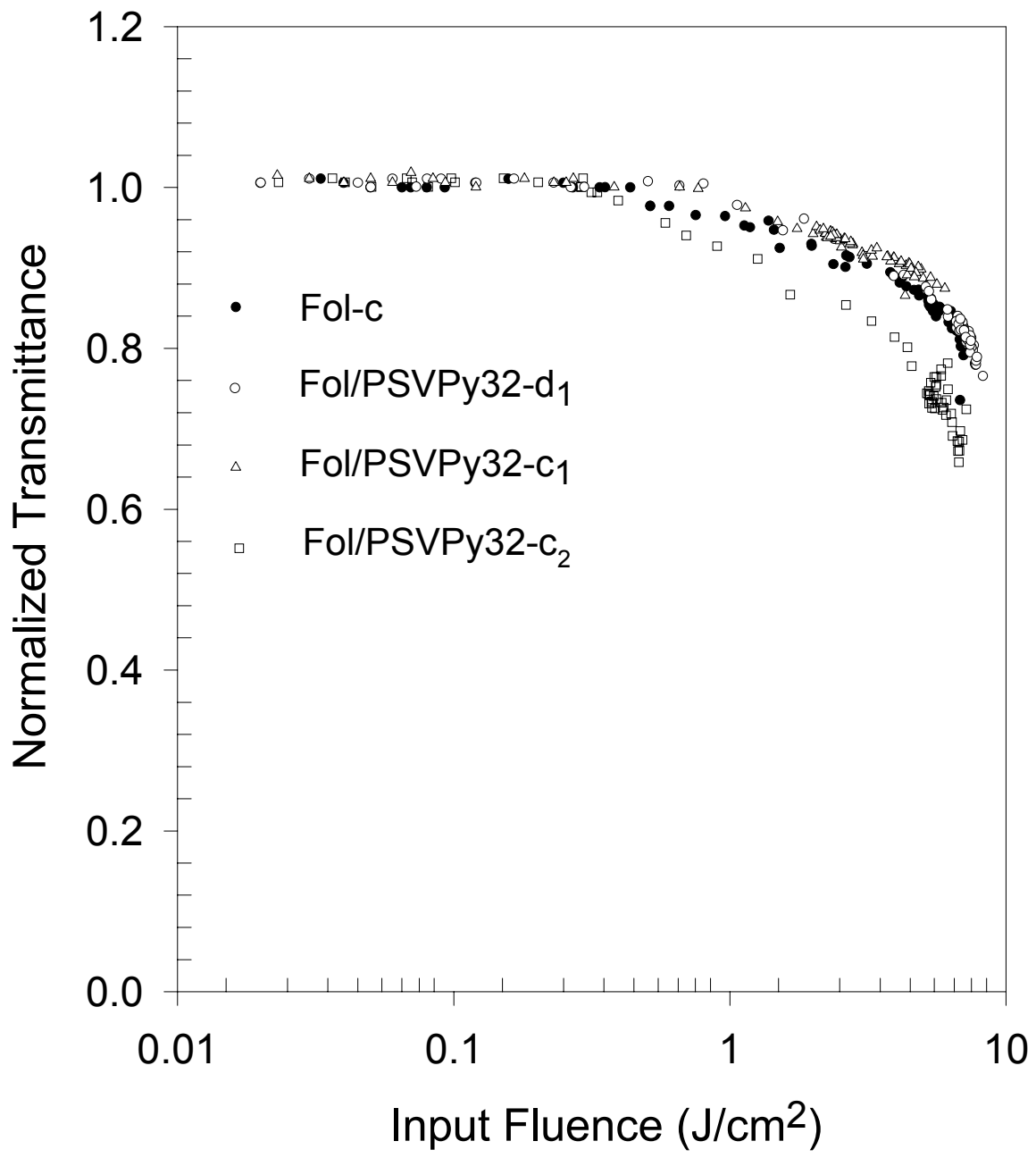

Fig. 7.

**"Nonlinear optics and optical limiting properties of multifunctional fullerenol/polymer composite"**

(Hendry Izaac Elim, *et al*)

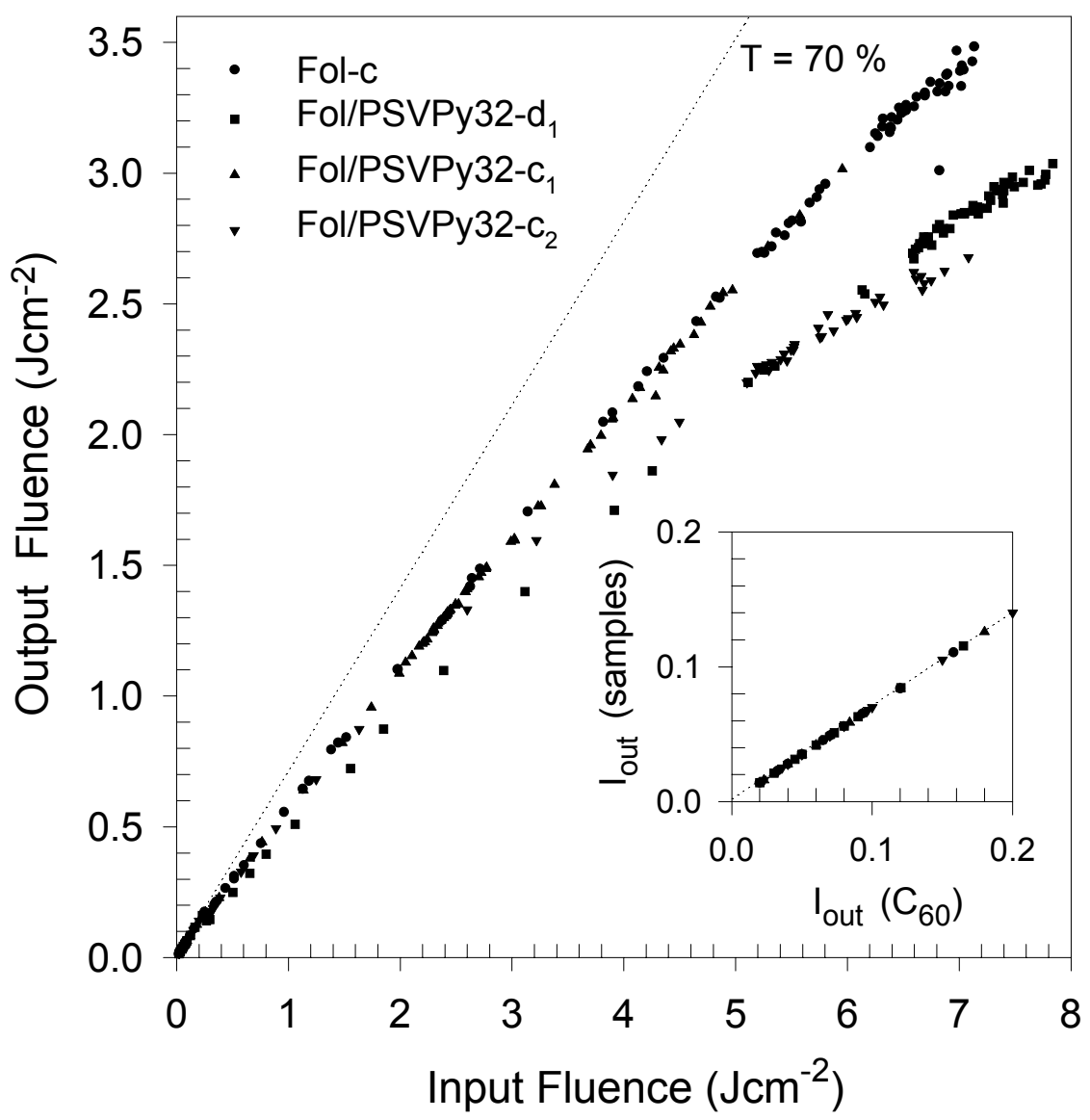

Fig. 8.

**"Nonlinear optics and optical limiting properties of multifunctional fullerenol/polymer composite"**

(Hendry Izaac Elim, *et al*)

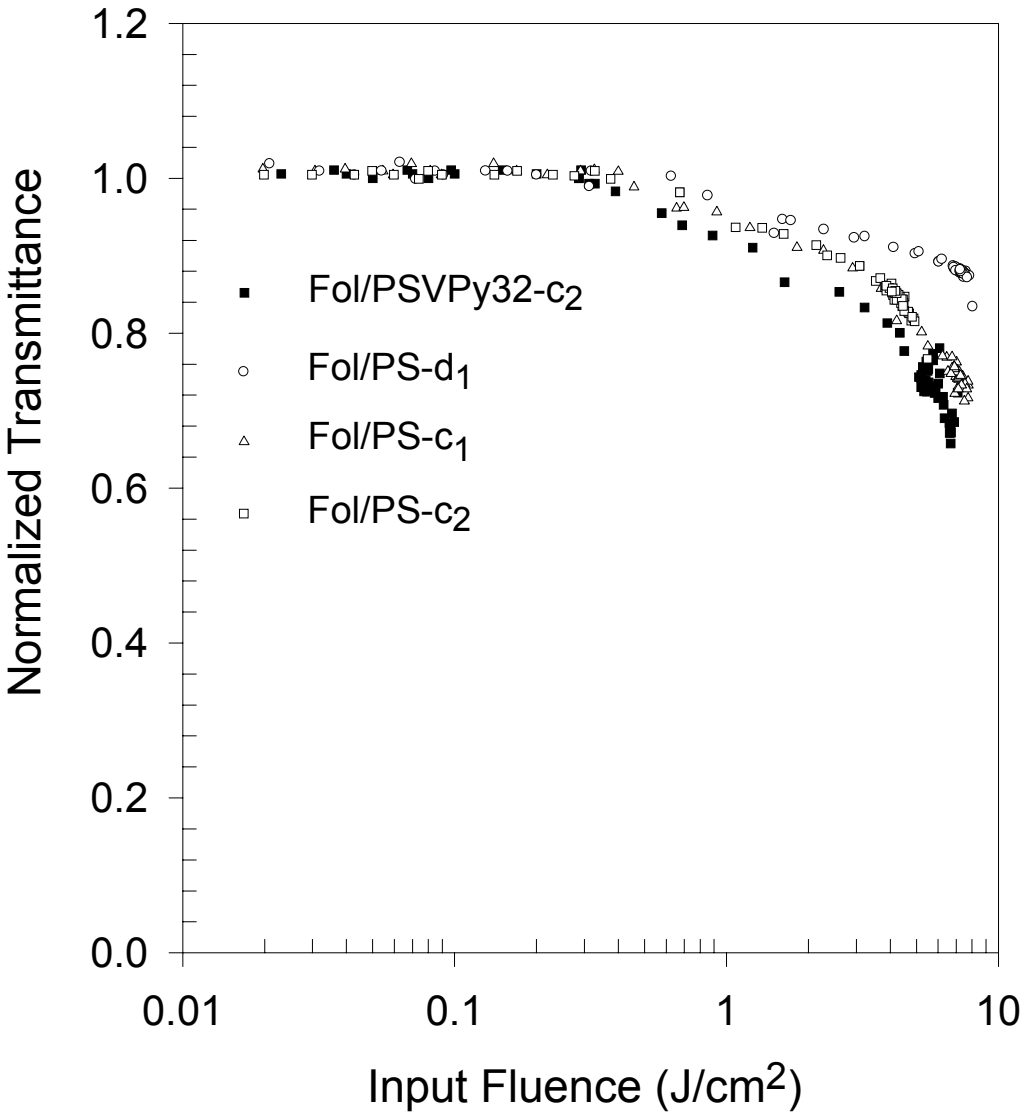

Fig. 9.

**"Nonlinear optics and optical limiting properties of multifunctional fullerenol/polymer composite"**

(Hendry Izaac Elim, *et al*)